\documentclass[sigconf]{acmart}

\newcommand{\wrt}{{{w.r.t.}}}
\AtBeginDocument{%
  \providecommand\BibTeX{{%
    \normalfont B\kern-0.5em{\scshape i\kern-0.25em b}\kern-0.8em\TeX}}}

\setcopyright{acmcopyright}
\copyrightyear{2018}
\acmYear{2018}
\acmDOI{10.1145/1122445.1122456}

\acmConference[Woodstock '18]{Woodstock '18: ACM Symposium on Neural
  Gaze Detection}{June 03--05, 2018}{Woodstock, NY}
\acmBooktitle{Woodstock '18: ACM Symposium on Neural Gaze Detection,
  June 03--05, 2018, Woodstock, NY}
\acmPrice{15.00}
\acmISBN{978-1-4503-XXXX-X/18/06}

\begin{document}

\title{CNN-based Local Vision Transformer for COVID-19 Diagnosis}

\author{Hongyan Xu}
\email{tjdxxhy@tju.edu.cn}
\affiliation{%
  \institution{the University of New South Wales}
  \streetaddress{111111}
  \city{Sydney}
  \country{Australia}
}

\author{Xiu Su}
\affiliation{%
  \institution{the University of Sydney}
  \city{Sydney}
  \country{Australia}}
\email{xisu5992@uni.sydney.edu.au}

\author{Dadong Wang}
\affiliation{%
  \institution{Data61, The Commonwealth Scientific and Industrial Research Organisation (CSIRO)}
  \streetaddress{1 Th{\o}rv{\"a}ld Circle}
  \city{Sydney}
  \country{Australia}}
\email{Dadong.wang@csiro.au}

\renewcommand{\shortauthors}{Hongyan Xu and Xiu Su, et al.}

\begin{abstract}
  Deep learning technology can be used as an assistive technology to help doctors quickly and accurately identify COVID-19 infections. Recently, Vision Transformer (ViT) has shown great potential towards image classification due to its global receptive field. However, due to the lack of inductive biases inherent to CNNs, the ViT-based structure leads to limited feature richness and difficulty in model training. In this paper, we propose a new structure called Transformer for COVID-19 (COVT) to improve the performance of ViT-based architectures on small COVID-19 datasets. It uses CNN as a feature extractor to effectively extract local structural information, and introduces average pooling to ViT's Multilayer Perception(MLP) module for global information. Experiments show the effectiveness of our method on the two COVID-19 datasets and the ImageNet dataset.
\end{abstract}




\maketitle

\section{Introduction}
As a global pandemic, the Coronavirus Disease 2019 (COVID-19) has so far spread to 192 countries and regions, causing 158,407,543 infections, resulting in the loss of 3,294,694 lives as of 10 May 2021\footnote{https://coronavirus.jhu.edu/map.html}. For the prevention and treatment of COVID-19, accurate identification of cases, isolation of patients from healthy people, and proper treatment of patients are of vital importance.

\begin{figure}[!t]
\centerline{\includegraphics[width=0.4\textwidth]{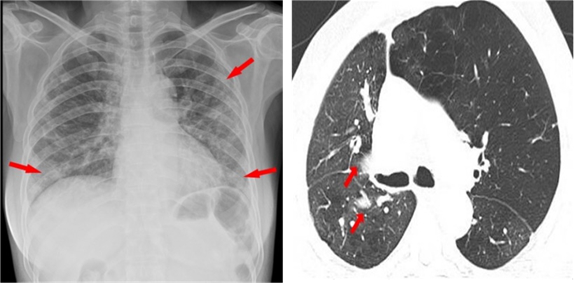}}
\caption{Examples of CXR(left) and CT(right) scans positive for COVID-19.  The red arrow in the picture indicates the area of COVID-19 infection.}
\label{fig4}
\vspace{-4mm}
\end{figure}

Real-time polymerase chain reaction (RT-PCR) is currently the most commonly used technique for COVID-19 detection. It detects respiratory samples and blood samples of subjects to be examined, and is known as the gold standard for COVID-19 detection due to its convenience and relatively high sensitivity [1]. However, the technology takes several hours from the beginning of the detection to obtaining the results. This delay in time may lead to contact between the patient and the normal population and cause the further spread of the virus [2]. Since the patient's computer tomography (CT) and chest X-ray imaging (CXR) images exhibit image characteristics shown in Fig. \ref{fig4}, such as ground-glass opacity (GGO), out-of-the-air bilateralities, and interstitiality [3-4], medical imaging processing will facilitate the rapid diagnosis of COVID-19.

However, in the early stage of COVID-19, its imaging findings are very similar to ordinary pneumonia and other lung diseases, and it is difficult to distinguish. In some underdeveloped areas, due to the lack of imaging professionals and related technical equipment, the problem will be more serious, causing a large number of patients to be misdetected or missed. Therefore, an efficient, convenient, and accurate covid-19 detection technology is needed. 

At present, some work that detects COVID-19 infection based on images has been proposed \cite{ouyang2020dual, soares2020sars,li2020efficient,li2020classification}. However, most of the existing methods have not been optimal due to insufficient labeled COVID-19 data, or are only validated on small datasets, and the model lacks generalization. Researchers have carried out a series of work to solve these problems, including the use of weakly supervised learning \cite{wang2020weakly}, anomaly detection \cite{2020anomaly}, and transfer learning \cite{2020Covid}. However, the effect is still limited because the models presented by these efforts lack generalization validation on large datasets.

Transformer structure was originally applied to natural language processing (NLP), is notable for its use of attention to model long-range dependencies in the data \cite{2021Swin}. Recently, it has performed well in computer vision tasks such as image classification \cite{2020An, 2020Training}, object detection \cite{2020End, 2020Deformable}, action
recognition \cite{2021Spatial}, joint vision-language modeling \cite{2021Learning},etc. However, due to the lack of inductive bias inherent to CNNs, it exhibits limited feature richness and difficulty in training when the training dataset is small. ViT \cite{2020An} point out that a hybrid ViT can be used, which uses CNN to generate initial feature embedding, and then train with Transformer. The introduction of the CNN structure can help improve the feature richness when the sample is insufficient, and re-introduces the inductive bias. 

\begin{figure*}[!t]
\centering
\includegraphics[width=0.8\linewidth]{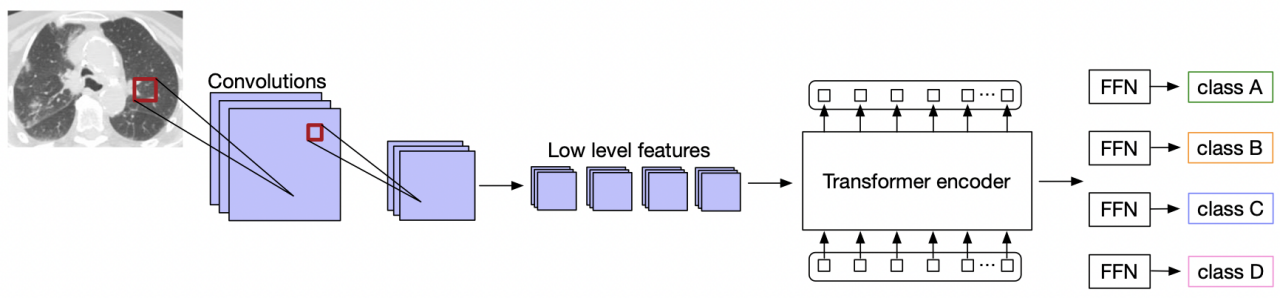}
\caption{The overall structure of COVT.}
\label{fig1}
\vspace{-4mm}
\end{figure*}

To solve the problem of model generalization in the COVID-19 detection task and the difficulty of training vision transformers on small datasets, we designed a hybrid model based on visual Transformer and CNN, called COVT, Transformer for COVID-19. This structure takes advantage of Transformer's ability to capture long-term dependence and CNN's ability to focus on local information and extract low-level information. Specifically, we designed an effective CNN backbone, which uses multiple atrous convolutions with different atrous rates to obtain multi-scale information without causing an increase in computational complexity and loss of image detail information. At the same time, in the Transformer structure, we have improved the MLP module to enable it to fuse global information and better capture long-term dependencies.

In summary, our contribution is as follows:

\begin{itemize}
    \item A multi-scale CNN backbone was designed to compensate for vision transformer's shortcomings in extracting low-level information.
    \item An improved MLP module is proposed to better integrate global information through average pooling.
	\item Experiments were performed on two COVID-19 CXR datasets. In order to further verify the generalization performance of the model on large dataset, experiments were conducted on the ImageNet dataset, and the results proved the validity of the proposed COVT model.
\end{itemize}

The rest of the paper is organized as follows. Section \ref{section2} reviews the related work. Section \ref{section3} introduces the proposed method. Section \ref{section4} describes the datasets used and the experiments conducted to evaluate the performance of the proposed method, followed by conclusion in Section \ref{section5}.

\begin{figure*}[!t]
\centering
\includegraphics[height =0.2\linewidth, width=0.8\linewidth]{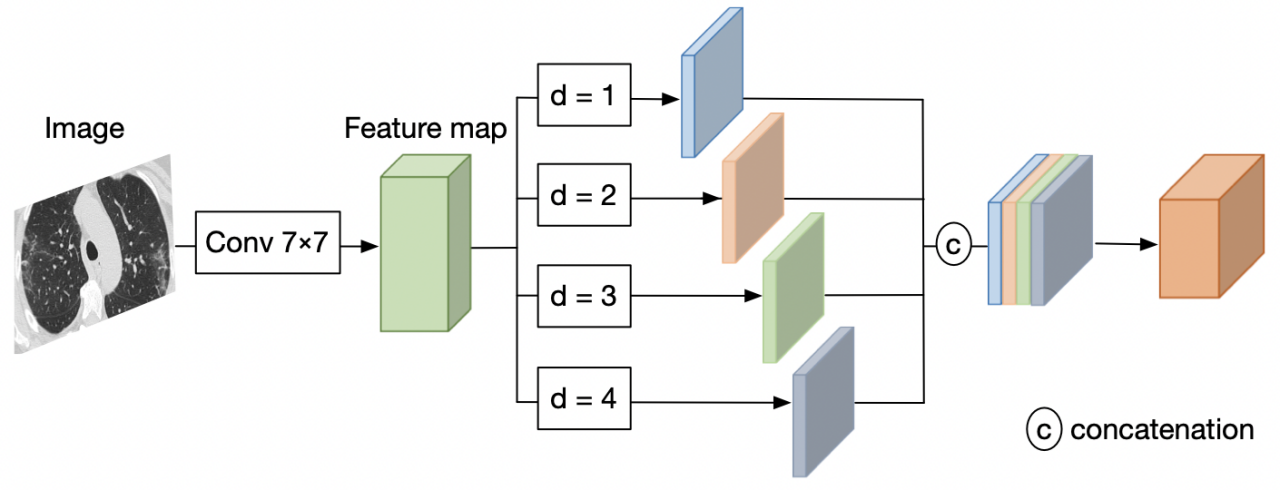}
\caption{The overall structure of CNN block.}
\label{fig2}
\vspace{-4mm}
\end{figure*}
\section{Related Work}\label{section2}
In this section, we discuss two types of works that are most relevant to our work, including deep learning based diagnosis of COVID-19 and Transformers in Vision.

\textbf{Deep Learning Based Diagnosis of COVID-19}
\qquad In recent years, convolutional neural networks(CNN) have been used to detect lung disease due to its effectiveness. Rajpurkar \textit{et al.} \cite{rajpurkar2018deep} designed a 121-layer network that performed well on the ChestX-ray14 \cite{wang2017chestx} dataset containing 14 different lung diseases. Gu \textit{et al.} \cite{gu2018classification} proposed a computer-aided diagnosis (CAD) system that can be used to identify bacterial and viral pneumonia in chest radiography. Inspired by these achievements, some scholars have applied deep learning-related algorithms to COVID-19 detection and achieved promising results. Tulin Ozturk \textit{et al.} \cite{0Automated} designed a model using raw chest X-ray images for automatic COVID-19 detection. The model achieved 98.08\% and 87.02\% accuracy on binary classification (COVID vs. No-Findings) and multi-class classification (COVID vs. No-Findings vs. Pneumonia) tasks, respectively. 

Eduardo Soares \textit{et al.} \cite{soares2020sars} established a public SARS-CoV-2 CT scan dataset, containing 1252 cases of SARS-CoV-2 infection (COVID-19) positive CT scans and 1,230 CT scans of not infected with SARS-CoV-2 but suffering from other lung diseases. Xuehai He \textit{et al.} \cite{he2020sample} used Self-Trans approach, which integrates self-supervised learning and transfer learning, to reach 0.85 F1 and 0.94 AUC when diagnosing COVID-19 using CT scans. However, due to the difficulty of COVID-19 data acquisition, most of these works are trained and tested on small datasets. None of these works verifies the generalization of the designed model on large-scale data sets. Although the generalization is very important when the model is used on different datasets.

\textbf{Transformer in Vision}
\qquad Transformer was first proposed by the Google team in 2017 for natural language processing (NLP) tasks, and then became the main backbone of NLP. In October 2020, ViT \cite{2020An}, the first pure Transformer applied in computer vision, was proposed, and achieved excellent results on large datasets such as JFT-300. However, ViT has shortcomings such as the inability to model local structures of images, such as edges and lines, and limited feature richness when training samples are insufficient, so a series of improvements have been made to the structure of ViT in studies such as \cite{2021Going, 2021Swin, 2021Incorporating}. MoCov3 \cite{2021moco} freezes the patch projection layer in ViT, i.e. uses a fixed random patch projection to alleviate the instability of the training process and improve model accuracy. DeiT \cite{2020Training} introduces knowledge distillation, which enables student models to learn useful information from teacher models through distillation token. T2T \cite{2021Tokens} recursively aggregating tokens into one token to improve the problem in ViT where simple tokenization of input images cannot model local information. Different from these works, this paper is dedicated to combining the advantages of CNN's inductive bias and local information extraction with the advantages of Transformer's long-term dependence. Combine local information and global information in an effective way to better guide image classification.

\section{Methods}\label{section3}
In this section, we describe the proposed Transformer for COVID-19 (COVT) in details.

\subsection{Network Structure}
In order to effectively extract local details and global context information from COVID-19 images, we designed a novel COVT framework, the overall structure of which is shown in Fig. \ref{fig1}.

The input COVID-19 image passes through the CNN block to extract multi-scale feature information. The low-level feature diagram is then entered into the Transform module to extract the global dependency. After passing the FFN, the results of the classification and the multi-scale receptive field and rich context information are output.
\subsection{CNN block}
When training from scratch on medium-sized datasets (such as ImageNet), ViT's performance is not as good as similar-sized CNN counterparts (e.g., ResNets). One of the reasons is that the direct tokenization of the input image through hard spilt makes ViT unable to model the local structure of the image (such as edges and lines), so it requires much more training samples than CNNs to achieve similar performance \cite{2021Tokens}. To alleviate this problem and improve the performance of ViT-based architectures on small COVID-19 datasets, we introduce a CNN block in the ViT structure.

 The designed CNN block is shown in Fig. \ref{fig2}. The input COVID-19 chest-X-ray(CXR) images first pass through a $7 \times 7$ convolution layer to increase the receptive field. This is because in deep neural networks, the size of the receptive field can roughly indicate the extent to which context information is used \cite{zhao2017pyramid}.  For CT and X-ray images, the COVID-19 infection area usually occupies only a small part of the image. Therefore, in order to accurately identify these areas, it is necessary to make effective use of contextual information and accurately extract image features. Thus, the feature map generated by the $7 \times 7$ convolution enters the spatial pyramid feature extraction module, and the multi-scale receptive field and rich context information are obtained through the atrous convolution with the atrous rates of 1, 2, 3, and 4 respectively. Then, the resulting feature maps, which are processed by four different atrous convolution branches, are concatenated to form a feature map that contains both rich local detail information and global information, and is input into the Transformer block.

\subsection{Transformer block}
\begin{figure}[!t]
\centering
\includegraphics[width=0.8\linewidth]{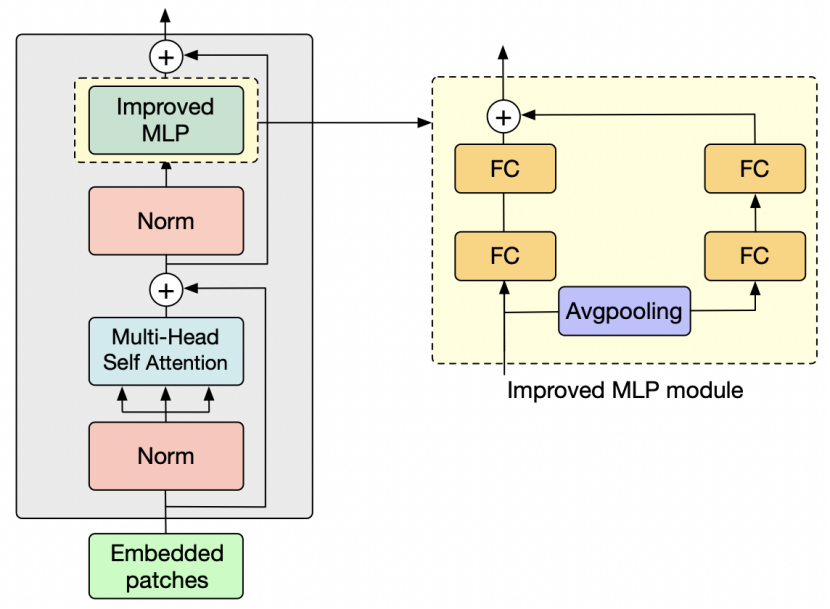}
\caption{The overall structure of Transformer block.}
\label{fig3}
\vspace{-4mm}
\end{figure}

The transformer block used in COVT is shown in Fig. \ref{fig3}. On the left is ViT's transformer encoder. We mainly improved the MLP module, which is referred to below as the Improved MLP module.

Suppose an input image token with a size of [B, N, C], where B is the batch size set during training, N means that the original feature map is divided into N patches, and C is the number of channels. For the MLP module in the Vision Transformer, it will be referred to as the original MLP hereinafter. When the image token is input to the MLP module for calculation, each patch inside the original MLP will be calculated separately, which is equivalent to using the local information inside the image token. In other words, the original MLP module focuses on local feature information. We made the improvements to the original MLP module as shown on the right side of Fig. \ref{fig3}, and the input image token will obtain global information through average pooling, at which point the size of the image token changes to [B,1,C]. After passing through two fully connected layers, For the output of the branch that has not gone through the average pooling layer, focuses on the local information of the image; for the output of the average pooling branch, it is rich in global information. After they are superimposed, it makes up for the shortcomings of the original MLP's insufficient attention to global information. At the same time, the added computational cost is very small and can be ignored.

\section{Experiments}\label{section4}

\textbf{Implementation Details.} \qquad In this section, we introduce the experimental details of the proposed method. In general, for all datasets \wrt~different models, we use SGD optimizer with momentum 0.9. The learning rate is decayed with cosine from 0.1 to $10^{-5}$ for all models. To accommodate the size \wrt~different datasets, we leverage batchsize of 64 for COVID X-ray dataset, 128 for covid5k dataset and ImageNet dataset. Besides, models are trained for 100 epochs before inference for all datasets. All experiments were performed on Nvidia Tesla V100 GPUs. The proposed model was constructed using Pytorch. 

\textbf{Datasets.} \qquad To verify the effectiveness of the proposed COVT model in detecting COVID-19 infection, we conducted verification on two COVID-19 datasets, the COVID\_xray dataset, and the covid5k dataset \cite{2020Deep}. The former contains 98 images, including 78 COVID-19 positive images and 20 COVID-19 undetected images. The latter, the covid5k dataset, contains 71 COVID-19 positive x-ray images and 5000 COVID-19 negative x-ray images. We also chose the ImageNet dataset as a validation dataset to test the generalization ability of our model. It contains 1.2M training images and 50K test images from 1,000 categories.
\subsection{Experiments on COVID\_xray dataset}
We conducted experiments on the COVID\_xray dataset. This is a tiny dataset with only 98 images. We use the deit-small and deit-tiny models as the backbone respectively. The COVT model using deit-small as the backbone is called COVT-S below, and the COVT model with deit-tiny as the backbone is called COVT-T. The effect of the COV-S and COV-T models on COVID\_xray datasets is shown in Table \ref{tabel111}.

 It can be seen that the proposed COVT model overcomes the disadvantages of the vision transformer's limited feature richness and difficulty in model training when the training samples are insufficient. The same effect as VGG16 is achieved on this dataset, with a detection accuracy of 100\%.
 
 \begin{table}[H]
\caption{Result of COVT and VGG16 on COVID\_xray dataset.}\label{tabel111}
\begin{tabular}{ccccc}
\hline
Model  & augmentation & Batch-size & Top-1 (\%) & Top-5 (\%) \\ \hline
VGG16  & /            & /          & 100        & 100        \\
COVT-S & mixup        & 64         & 100        & 100        \\
COVT-T & nomixup      & 64         & 100        & 100        \\ \hline
\end{tabular}
\end{table}

\subsection{Experiments on covid5k dataset}
To further verify the effectiveness of the COVT model in detecting COVID-19 infection, we conducted experiments on the covid5k dataset. Table \ref{tabel222} lists the experimental results of the COVT model and its comparison with several CNN models on the dataset.

The accuracy of our COVT-S and COVT-T models on the covid5k dataset is equivalent to or better than the convolutional neural network     in Table \ref{tabel222}. It is worth noting that the COVT-T model is 0.06\% better than the COVT-S model and 0.04\% better than the GDCNN model. Compared with the other two convolutional neural networks, COVT shows greater superiority. The accuracy of the COVT-S model is 11.40\% and 3.85\% higher than that of ADCNN and ResNet50, respectively, and the COVT-T model is 11.46\% and 3.91\%, respectively.

\begin{table}[H]
\caption{Result of COVT and other networks on covid5k dataset.}\label{tabel222}
\setlength{\tabcolsep}{1.2mm}{
\begin{tabular}{ccccc}
\hline
Model    & augmentation & Batch-size & \begin{tabular}[c]{@{}c@{}}Top-1 (\%)\end{tabular} & \multicolumn{1}{l}{Top-5 (\%)} \\ \hline
GDCNN \cite{2020Prediction}    & /            & /          & 98.84                                                & /                              \\
ACNN \cite{2020Weighted}     & /            & /          & 87.42                                                & /                              \\
ResNet50 \cite{2020Weighted} & /            & /          & 94.97                                                & /                              \\
COVT-S   & mixup        & 128        & 98.82                                                & 100                            \\
COVT-T   & nomixup      & 128        & 98.88                                                & 100                            \\ \hline
\end{tabular}}
\end{table}
\subsection{Experiments on ImageNet dataset}
 We use the same training settings as DeiT to train the COVT-S and COVT-T models. To demonstrate the effectiveness of the COVT model relative to CNN, we also included representative CNN models such as ResNet and GhostNet \cite{2020GhostNet} in the comparison. It can be seen that our COVT-S model is better than other transformer-based models in Table \ref{tabel333}. In particular, the COVT-T model achieves 76.15\% top-1 accuracy, which is 3.95\% higher than the baseline model DeiT-tiny,    which shows the benefits of introducing the COVT framework to retain local structural information and enhance global dependencies. Our COVT-S model is better than the baseline model DeiT-small 0.83\% and 2.73\% higher than the ViT-B/16 model. Compared with several CNN models listed in the table, COVT-S is better than the widely used ResNet50 and ResNet152 models and 1.23\% more accurate than the GhostNet-A model. The effect of the COVT-T model is lower than that of the ResNet152 and GhostNet-A models, but it is 0.43\% higher than the ResNet50 model. Given the computational complexity of the COVT-T model, this effect is acceptable.
 
 \begin{table}[H]
\caption{Result of COVT and other networks on ImageNet dataset.}\label{tabel333}
\begin{tabular}{cccc}
\hline
Model             & Resolution & \multicolumn{1}{l}{Top-1(\%)} & Top-5 (\%) \\ \hline
CNN-based         &            &                               &            \\
ResNet50          & $224\times224$    & 76.20                       & 92.90    \\
ResNet152         & $224\times224$    & 78.30                       & 94.10    \\
GhostNet-A \cite{2020GhostNet}        & $224\times224$    & 79.40                       & 94.50    \\ \hline
Transformer-based &            &                               &            \\
DeiT-small \cite{2020Training}        & $224\times224$    & 79.80                       & /          \\
DeiT-tiny \cite{2020Training}         & $224\times224$    & 72.20                       & 91.10    \\
COVT-S            & $224\times224$    & 80.63                       & 94.83    \\
COVT-T            & $224\times224$    & 76.15                       & 92.59    \\
ViT-B/16 \cite{2020An}         & $384\times384$    & 77.90                       & /          \\
ViT-L/16 \cite{2020An}          & $384\times384$           & 76.50                       & /          \\ \hline
\end{tabular}
\end{table}

\section{Conclusion}\label{section5}
In this paper, We propose a general method that can be integrated into common classification networks to improve their performance in COVID-19 infection detection task. In detail, our D-SPP module can be used to collect multi-scale image features and context information, and provide guidance for subsequent accurate predictions. Our proposed CID module can maintain the focus of the CNN network on areas of interest related to COVID- 19 infection. In addition, the proposed modules can be easily integrated into various deep learning networks to improve their performance. Extensive experiments have been conducted on four COVID-19 CT and CXR image datasets to evaluate the performance of the proposed method, and the experimental results show the superiority of our approach to other state-of-the-art methods.

\begin{acks}
The authors would like to thank the creators of all COVID-19 datasets used in this paper for making the datasets publicly available.
\end{acks}


\newpage
\bibliographystyle{ACM-Reference-Format}
\bibliography{sample-base}

\begin{thebibliography}{10}
\providecommand{\url}[1]{#1}
\csname url@samestyle\endcsname
\providecommand{\newblock}{\relax}
\providecommand{\bibinfo}[2]{#2}
\providecommand{\BIBentrySTDinterwordspacing}{\spaceskip=0pt\relax}
\providecommand{\BIBentryALTinterwordstretchfactor}{4}
\providecommand{\BIBentryALTinterwordspacing}{\spaceskip=\fontdimen2\font plus
\BIBentryALTinterwordstretchfactor\fontdimen3\font minus
  \fontdimen4\font\relax}
\providecommand{\BIBforeignlanguage}[2]{{%
\expandafter\ifx\csname l@#1\endcsname\relax
\typeout{** WARNING: IEEEtran.bst: No hyphenation pattern has been}%
\typeout{** loaded for the language `#1'. Using the pattern for}%
\typeout{** the default language instead.}%
\else
\language=\csname l@#1\endcsname
\fi
#2}}
\providecommand{\BIBdecl}{\relax}
\BIBdecl

\bibitem{2020Covid}
I.~D. Apostolopoulos and T.~Bessiana, ``Covid-19: automatic detection from
  x-ray images utilizing transfer learning with convolutional neural
  networks,'' 2020.

\bibitem{2020Prediction}
R.~G. Babukarthik, V.~Adiga, G.~Sambasivam, D.~Chandramohan, and J.~Amudhavel,
  ``Prediction of covid-19 using genetic deep learning convolutional neural
  network (gdcnn),'' \emph{IEEE Access}, vol.~8, pp. 177\,647--177\,666, 2020.

\bibitem{2020End}
N.~Carion, F.~Massa, G.~Synnaeve, N.~Usunier, A.~Kirillov, and S.~Zagoruyko,
  ``End-to-end object detection with transformers,'' 2020.

\bibitem{2021moco}
X.~Chen, S.~Xie, and K.~He, ``An empirical study of training self-supervised
  visual transformers,'' 04 2021.

\bibitem{2020An}
A.~Dosovitskiy, L.~Beyer, A.~Kolesnikov, D.~Weissenborn, and N.~Houlsby, ``An
  image is worth 16x16 words: Transformers for image recognition at scale,''
  2020.

\bibitem{gu2018classification}
X.~Gu, L.~Pan, H.~Liang, and R.~Yang, ``Classification of bacterial and viral
  childhood pneumonia using deep learning in chest radiography,'' in
  \emph{Proceedings of the 3rd International Conference on Multimedia and Image
  Processing}, 2018, pp. 88--93.

\bibitem{2020GhostNet}
K.~Han, Y.~Wang, Q.~Tian, J.~Guo, and C.~Xu, ``Ghostnet: More features from
  cheap operations,'' in \emph{2020 IEEE/CVF Conference on Computer Vision and
  Pattern Recognition (CVPR)}, 2020.

\bibitem{he2020sample}
X.~He, X.~Yang, S.~Zhang, J.~Zhao, Y.~Zhang, E.~Xing, and P.~Xie,
  ``Sample-efficient deep learning for covid-19 diagnosis based on ct scans,''
  \emph{medRxiv}, 2020.

\bibitem{li2020classification}
C.~Li, D.~Dong, L.~Li, W.~Gong, X.~Li, Y.~Bai, M.~Wang, Z.~Hu, Y.~Zha, and
  J.~Tian, ``Classification of severe and critical covid-19 using deep learning
  and radiomics,'' \emph{IEEE Journal of Biomedical and Health Informatics},
  vol.~24, no.~12, pp. 3585--3594, 2020.

\bibitem{li2020efficient}
Y.~Li, D.~Wei, J.~Chen, S.~Cao, H.~Zhou, Y.~Zhu, J.~Wu, L.~Lan, W.~Sun, T.~Qian
  \emph{et~al.}, ``Efficient and effective training of covid-19 classification
  networks with self-supervised dual-track learning to rank,'' \emph{IEEE
  Journal of Biomedical and Health Informatics}, vol.~24, no.~10, pp.
  2787--2797, 2020.

\bibitem{2021Swin}
Z.~Liu, Y.~Lin, Y.~Cao, H.~Hu, Y.~Wei, Z.~Zhang, S.~Lin, and B.~Guo, ``Swin
  transformer: Hierarchical vision transformer using shifted windows,'' 2021.

\bibitem{2020Deep}
S.~Minaee, R.~Kafieh, M.~Sonka, S.~Yazdani, and G.~J. Soufi, ``Deep-covid:
  Predicting covid-19 from chest x-ray images using deep transfer learning,''
  \emph{Medical Image Analysis}, vol.~65, 2020.

\bibitem{ouyang2020dual}
X.~Ouyang, J.~Huo, L.~Xia, F.~Shan, J.~Liu, Z.~Mo, F.~Yan, Z.~Ding, Q.~Yang,
  B.~Song \emph{et~al.}, ``Dual-sampling attention network for diagnosis of
  covid-19 from community acquired pneumonia,'' \emph{IEEE Transactions on
  Medical Imaging}, 2020.

\bibitem{2020Weighted}
{\"O}.~{\"O}zdemir and E.~B. S{\"o}nmez, ``Weighted cross-entropy for
  unbalanced data with application on covid x-ray images,'' in \emph{2020
  Innovations in Intelligent Systems and Applications Conference (ASYU)}.\hskip
  1em plus 0.5em minus 0.4em\relax IEEE, 2020, pp. 1--6.

\bibitem{2021Spatial}
C.~Plizzari, M.~Cannici, and M.~Matteucci, \emph{Spatial Temporal Transformer
  Network for Skeleton-Based Action Recognition}.\hskip 1em plus 0.5em minus
  0.4em\relax Pattern Recognition. ICPR International Workshops and Challenges,
  2021.

\bibitem{2021Learning}
A.~Radford, J.~W. Kim, C.~Hallacy, A.~Ramesh, G.~Goh, S.~Agarwal, G.~Sastry,
  A.~Askell, P.~Mishkin, and J.~Clark, ``Learning transferable visual models
  from natural language supervision,'' 2021.

\bibitem{rajpurkar2018deep}
P.~Rajpurkar, J.~Irvin, R.~L. Ball, K.~Zhu, B.~Yang, H.~Mehta, T.~Duan,
  D.~Ding, A.~Bagul, C.~P. Langlotz \emph{et~al.}, ``Deep learning for chest
  radiograph diagnosis: A retrospective comparison of the chexnext algorithm to
  practicing radiologists,'' \emph{PLoS medicine}, vol.~15, no.~11, p.
  e1002686, 2018.

\bibitem{soares2020sars}
E.~Soares, P.~Angelov, S.~Biaso, M.~H. Froes, and D.~K. Abe, ``Sars-cov-2
  ct-scan dataset: A large dataset of real patients ct scans for sars-cov-2
  identification,'' \emph{medRxiv}, 2020.

\bibitem{0Automated}
A.~To, B.~Mt, C.~Eay, D.~Ubb, E.~Oy, and H.~Urafg, ``Automated detection of
  covid-19 cases using deep neural networks with x-ray images,''
  \emph{Computers in Biology and Medicine}, vol. 121.

\bibitem{2021Going}
H.~Touvron, M.~Cord, A.~Sablayrolles, G.~Synnaeve, and H.~Jégou, ``Going
  deeper with image transformers,'' 2021.

\bibitem{2020Training}
H.~Touvron, M.~Cord, M.~Douze, F.~Massa, and H.~Jégou, ``Training
  data-efficient image transformers distillation through attention,'' 2020.

\bibitem{wang2020weakly}
X.~Wang, X.~Deng, Q.~Fu, Q.~Zhou, J.~Feng, H.~Ma, W.~Liu, and C.~Zheng, ``A
  weakly-supervised framework for covid-19 classification and lesion
  localization from chest ct,'' \emph{IEEE Transactions on Medical Imaging},
  2020.

\bibitem{wang2017chestx}
X.~Wang, Y.~Peng, L.~Lu, Z.~Lu, M.~Bagheri, and R.~M. Summers, ``Chestx-ray8:
  Hospital-scale chest x-ray database and benchmarks on weakly-supervised
  classification and localization of common thorax diseases,'' in
  \emph{Proceedings of the IEEE conference on computer vision and pattern
  recognition}, 2017, pp. 2097--2106.

\bibitem{2021Incorporating}
K.~Yuan, S.~Guo, Z.~Liu, A.~Zhou, F.~Yu, and W.~Wu, ``Incorporating convolution
  designs into visual transformers,'' 2021.

\bibitem{2021Tokens}
L.~Yuan, Y.~Chen, T.~Wang, W.~Yu, Y.~Shi, F.~E. Tay, J.~Feng, and S.~Yan,
  ``Tokens-to-token vit: Training vision transformers from scratch on
  imagenet,'' 2021.

\bibitem{2020anomaly}
J.~Zhang, Y.~Xie, Y.~Li, C.~Shen, and Y.~Xia, ``Covid-19 screening on chest
  x-ray images using deep learning based anomaly detection,'' 03 2020.

\bibitem{zhao2017pyramid}
H.~Zhao, J.~Shi, X.~Qi, X.~Wang, and J.~Jia, ``Pyramid scene parsing network,''
  in \emph{Proceedings of the IEEE conference on computer vision and pattern
  recognition}, 2017, pp. 2881--2890.

\bibitem{2020Deformable}
X.~Zhu, W.~Su, L.~Lu, B.~Li, and J.~Dai, ``Deformable detr: Deformable
  transformers for end-to-end object detection,'' 2020.

\end{thebibliography}
\end{document}